\documentclass[12pt,a4paper]{article}
\usepackage{epsfig}
\begin{document}
\title{Littlest Higgs model and top-charm production at high-energy linear colliders}
\author{Chong-Xing Yue, Lei Wang, Yi-Qun Di, Shuo Yang\\
{\small  Department of Physics, Liaoning Normal University, Dalian
116029, China}\thanks{E-mail:cxyue@lnnu.edu.cn}\\}
\date{\today}

\maketitle
\begin{abstract}
Due to the presence of extra top quark T in the little Higgs
models, the $CKM$ matrix is not unitary and the flavor changing
neutral currents may exist at the tree level. In the context of
the Littlest Higgs(LH) model, we discuss the top-charm production
at the high-energy linear $e^{+}e^{-}$ collider $(LC)$ via the
processes $e^{+}e^{-}\rightarrow \overline{t}c +t\overline{c}$,
$e^{+}e^{-}\rightarrow (\overline{t}c
+t\overline{c})\nu_{e}\overline{\nu_{e}}$, and
$e^{-}\gamma\rightarrow e^{-}\overline{t}c$. We find that the
resonance production cross section for the process
$e^{+}e^{-}\rightarrow \overline{t}c+ t\overline{c} $ is
significantly larger, which can be detected in future $LC$
experiments.

 \vspace{1cm}

PACS number: 12.60.Cn, 14.70.Pw, 14.80.Cp

\end{abstract}

\newpage
\noindent{\bf 1. Introduction}

It is well known that, in the standard model$(SM)$, there is no
flavor changing neutral currents($FCNC's$) at tree-level and at
one-loop level they are $GIM$ suppressed. Searching for $FCNC's$
is one of the most interesting means to test the $SM$ and probe
popular new physics models. The top quark, with a mass of the
order of the electroweak scale $m_{t}=178.0\pm4.36 GeV$[1], is the
heaviest particle yet discovered. In some new physics models, the
$FCNC$ couplings involving the top quark may be significantly
enhanced[2]. Thus, searching for $FCNC's$ involving the top quark
would be a good probe for new physics beyond the $SM$.

The top quark $FCNC$ processes can be studied either in the rare
top quark decays or in the top quark production through $FCNC$
couplings at high-energy experiments. In the $SM$, such kind of
processes are unobservably small. Any signal of these processes
will be a clear evidence of new physics beyond the $SM$. Many new
physics models predict the existence of the $FCNC$ coupling
vertices $tcv(v=Z, \gamma,$ $or$ $g$), which can enhance the
branching ratios $Br(t\rightarrow cv)$ and the cross sections of
the top-charm production processes by several orders to make them
potentially accessible at future high energy collider
experiments[2]. The top-charm production processes have been
extensively studied in the context of some specific popular
models[3,4,5] and in a model independent approach[6]. They have
shown that some of new physics models might be tested or be
constrained through studying their effects on the top-charm
production processes.

To solve the so-called hierarchy or fine-tuning problem of the
$SM$, the little Higgs theory[7] was proposed as kind of
electroweak symmetry breaking ($EWSB$) mechanism accomplished by a
naturally light Higgs sector. This kind of models provide a
natural mechanism of $EWSB$ associated with the large value of the
top quark Yukawa couplings. This mechanism typically involves a
new heavy $SU(2)_{L}$ single top quark $T$. The existence of the
vector-like top quark $T$ introduces new effects in the weak
currents. The $CKM$ matrix is extended to $4\times3$ and $FCNC's$
occur at tree-level[8,9]. It has been shown[9] that the flavor
change $Z$ couplings are allowed in the up quark sector but not in
the down quark sector, which might be tested via rare top decays
and same sign top pair production at the $LHC$ experiments. In
this Letter, we will study the top-charm production induced by the
littlest Higgs $(LH)$ model[10] at the future high-energy linear
$e^{+}e^{-}$ collider $(LC)$ experiments and see whether the $FC$
signals of the $LH$ model can be detected via the top-charm
production.

The FC couplings $Z\overline{t}c$ and $Z_{H}\overline{t}c$ induced
by the vector-like top quark $T$ in the $LH$ model are given in
section 1. The contributions of these $FC$ couplings to the
process $e^{+}e^{-}\rightarrow \overline{t}c+t\overline{c}$ are
also calculated in this section. The contributions of these
couplings to the t-channel vector boson fusion processes
$e^{+}e^{-}\rightarrow W^{*}W^{*}
\overline{\nu}_{e}\nu_{e}\rightarrow (\overline{t}c +t\overline{c}
)\overline{\nu}_{e}\nu_{e}$ and $e^{-}\gamma\rightarrow
e^{-}\overline{t}c$ are further calculated in section 3 and 4,
respectively. Our conclusions are given in section 5.

\noindent{\bf 2. The process $e^{+}e^{-}\rightarrow
\overline{t}c+t\overline{c}$ in the $LH$ model}

It is well known that the most dangerous radiative corrections to
the Higgs mass in the $SM$ come from one-loop diagrams with top
quark, $SU(2)$ gauge bosons, and the Higgs self-coupling. In the
little theory[7], the Higgs mass is protected from one-loop
quadratic divergences by approximate global symmetries. New
particles, such as heavy scalars, heavy fermions and gauge bosons,
must be introduced to ensure that the global symmetries are not
broken too severely and to cancel the one-loop quadratic
divergence of the Higgs mass-squared. Furthermore, the numerically
most large quadratic divergence comes from top quark loops. The
cancellation of the quadratic divergence associated with the top
Yukawa coupling is the most important. Thus, all of the little
Higgs models should predict the existence of at least one
vector-like top quark at the $TeV$ scale.

In general, the presence of a new extra quark modifies the
electroweak currents. In the $LH$ model, the new vector-like top
quark $T$ makes that the number of up-type quarks is four and the
$3\times3$ $CKM$ matrix in the $SM$, which is related the quark
mass eigen states with the weak eigen states, became to a
$4\times3$ matrix. Since the top quark t and the vector-like quark
$T$ have different $SU(2)\otimes U(1)$ quantum numbers, their
mixing can lead to the $FCNC's$ mediated by the $SM$ gauge boson
$Z$. In the $LH$ model, the $FC$ couplings involving the top quark
can be written as[8,9]:
\begin{eqnarray}
\pounds\nonumber&=&\frac{e}{2S_{W}C_{W}}(K_{tu}\overline{t_{L}}\gamma_{\mu}u_{L}
+K_{tc}\overline{t_{L}}\gamma_{\mu}c_{L})Z^{\mu}\\
&&+\frac{e}{2S_{W}C_{W}}
\frac{c}{s}(K_{tu}\overline{t_{L}}\gamma_{\mu}u_{L}
+K_{tc}\overline{t_{L}}\gamma_{\mu}c_{L})Z^{\mu}_{H}+h.c. ,
\end{eqnarray}
where $S_{W}=\sin\theta_{W}$, $\theta_{W}$ is the Weinberg angle,
$c(s=\sqrt{1-c^{2}})$ is the mixing parameter between $SU(2)_{1}$
and $SU(2)_{2}$ gauge bosons. $Z_{H}$ is the new $SU(2)$ gauge
boson predicted by the $LH$ model. The factors $K_{tc}$ and
$K_{tu}$ are the off-diagonal matrix elements of the $4 \times 4$
neutral currents mixing matrix in the up-type quark sector, which
comes from the up-type quark transformation matrix. Reference[9]
has estimated the values of these factors via considering a
perturbative diagonalization of the up-type quark mass matrix and
found that their values are approximately equal to
$2.43\times10^{-3}$ and $2.12\times10^{-4}$, respectively.

In the $LH$ model, the $FC$ process $e^{+}e^{-}\rightarrow
\overline{t}c+t\overline{c}$ can be generated by the tree-level
$FC$ couplings $Z\overline{t}c$ and $Z_{H}\overline{t}c$. The
total cross section of this process can be written as:
\begin{eqnarray}
\sigma(S)\nonumber&=&\sigma_{Z}+\sigma_{Z_{H}}+\sigma_{ZZ_{H}}\\\nonumber
&=&\frac{\pi\alpha^{2}_{e}K^{2}_{tc}}
{4S^{4}_{W}C^{4}_{W}}\{(1-4S^{4}_{W}+8S^{4}_{W})\beta^{4}(3-\beta^{2})S
\chi^{2}_{Z}+\frac{C^{2}_{W}c^{2}}{s^{2}}\beta^{4}(3-\beta^{2})S
\chi^{2}_{ZH}\\&&
+\frac{8C_{W}c}{s}(1-2S^{2}_{W})\beta^{4}(3-\beta^{2})R_{e}[\chi_{Z}\cdot\chi_{Z_{H}}]\}
\end{eqnarray}
with
\begin{equation}
\chi_{i}=\frac{1}{S-M^{2}_{i}+iM_{i}\Gamma_{i}},
\end{equation}
where $\Gamma_{i}(i=Z,$ or $Z_{H})$ represents the total decay
width of the gauge bosons $Z$ or $Z_{H}$. $S$ is the
center-of-mass($CM$) energy squared.

From above equations, we can see that the cross section
$\sigma(S)$ of the top-charm production via the process
$e^{+}e^{-}\rightarrow \overline{t}c+t\overline{c}$ mainly depends
on the free parameters $c$ and $M_{Z_{H}}$ for the fixed value of
the flavor factor $K_{tc}$. Taking into account the precision
electroweak constrains on the parameter space of the $LH$ model,
the free parameters $c$ and the $Z_{H}$ mass $M_{Z_{H}}$ are
allowed in the ranges of $0\sim0.5$ and $1\sim2TeV$[11]. If we
take the  $CM$ energy $\sqrt{S}=500GeV$, then there is $S\ll
M^{2}_{Z_{H}}$. In this case, the contributions of the $LH$ model
to the top-charm production via the process $e^{+}e^{-}\rightarrow
\overline{t}c+t\overline{c}$ at a $LC$ with $\sqrt{S}=500GeV$
mainly come from the $FC$ coupling $Z\overline{t}c$. The value of
the cross section $\sigma(S)$ is not sensitive to the free
parameters $c$ and $M_{Z_{H}}$, and is about $1.22\times10^{-2}fb$
in most of all parameter space preferred by the electroweak
precision data. Comparing with the $SM$ prediction, the cross
section $\sigma(S)$ is enhanced by several orders of magnitude.
However, there will be only several $\overline{t}c$ events to be
generated in the future $LC$ experiment with $\sqrt{S}=500GeV$ and
a yearly integrated luminosity of $\pounds=340fb^{-1}$[12], which
is very difficult to be detected.

\begin{figure}[htb]
\vspace{-0.5cm}
\begin{center}
\epsfig{file=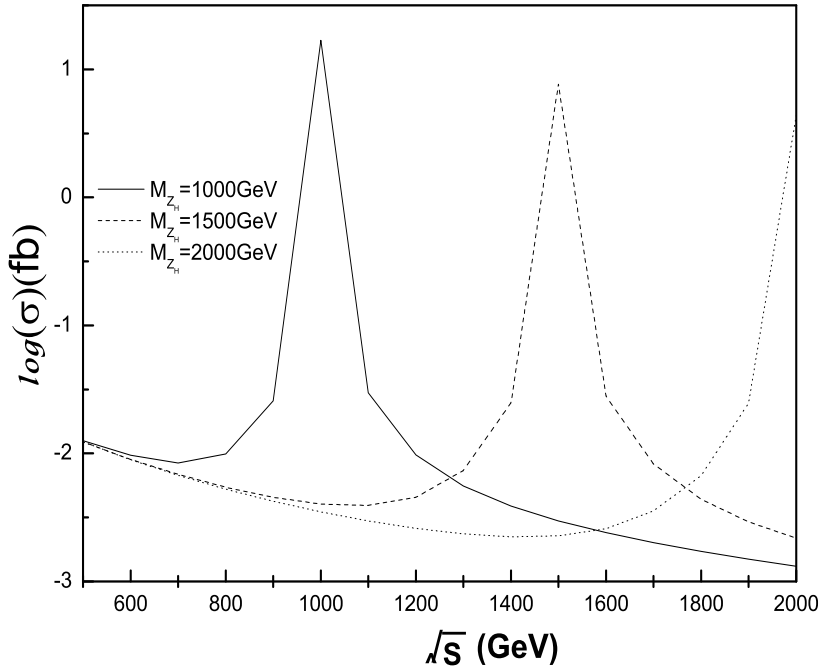,width=350pt,height=300pt} \vspace{-1.0cm}
\hspace{5mm} \caption{The top-charm production cross section
$\sigma(S)$ as a function of the $CM$ energy
\hspace*{1.8cm}$\sqrt{S}$ for three values of $M_{Z_{H}}$.}
\label{ee}
\end{center}
\end{figure}
To see the effects of the $CM$ energy $\sqrt{S}$ on the top-charm
production, we plot the $\sigma(S)$ versus $\sqrt{S}$ in Fig.1 for
$c=0.4$ and three values of $M_{Z_{H}}$. From Fig.1, we can see
that the cross section $\sigma(S)$ resonance emerges when the
$Z_{H}$ mass $M_{Z_{H}}$ approaches the $CM$ energy $\sqrt{S}$. In
this case, the contributions of the $LH$ model to the top-charm
production mainly come from the $FC$ coupling
$Z_{H}\overline{t}c$. The resonance values of the $\sigma(S)$
decrease as $\sqrt{S}$ increasing. For $c=0.4$ and
$\sqrt{S}=M_{Z_{H}}=1TeV, 1.5TeV$, and $2TeV$, the cross section
$\sigma(S)$ can reach $16.8fb, 7.7fb$, and $4.4fb$, respectively.
Then there will be several hundreds and up to thousands
$\overline{t}c$ events to be generated at the future $LC$
experiments with $\pounds=500fb^{-1}$ and $\sqrt{S}\geq 1TeV$,
which should be observable. Thus, the possible $FC$ signals of the
$LH$ model can be detected in future $LC$ experiments.

\noindent{\bf 3. The process $e^{+}e^{-}\rightarrow WW
\nu_{e}\overline{\nu_{e}}\rightarrow (\overline{t}c
+t\overline{c})\nu_{e} \overline{\nu_{e}}$ in the $LH$ model}

The $WW$-fusion process $e^{+}e^{-}\rightarrow WW
\nu_{e}\overline{\nu_{e}}\rightarrow(\overline{t}c +
t\overline{c}) \nu_{e}\overline{\nu}_{e}$ is very sensitive to the
$FC$ couplings[5,6]. Thus, the $FC$ couplings $Z\overline{t}c$ and
$Z_{H}\overline{t}c$ might be probed via this process in future
$LC$ experiments. Furthermore, the cross section of this process
grows with the $CM$ energy $\sqrt{S}$ of the $LC$ experiment,
while the production cross section of the $s$-channel process
$e^{+}e^{-}\rightarrow \overline{t}c + t\overline{c}$ generally
drops as $\sqrt{S}$ increasing. Thus, there is a strong motivation
to study the $WW$ process $e^{+}e^{-}\rightarrow WW
\nu_{e}\overline{\nu}_{e}\rightarrow(\overline{t}c +
t\overline{c})\nu_{e}\overline{\nu_{e}}$ at somewhat higher $CM$
energies. In this section, we consider the contributions of the
$FC$ couplings $Z\overline{t}c$ and $Z_{H}\overline{t}c$ to this
process in the context of the $LH$ model, the relevant Feynman
diagrams are shown in Fig.2.

\begin{figure}[htb]
\vspace{-8.5cm}
\begin{center}
\epsfig{file=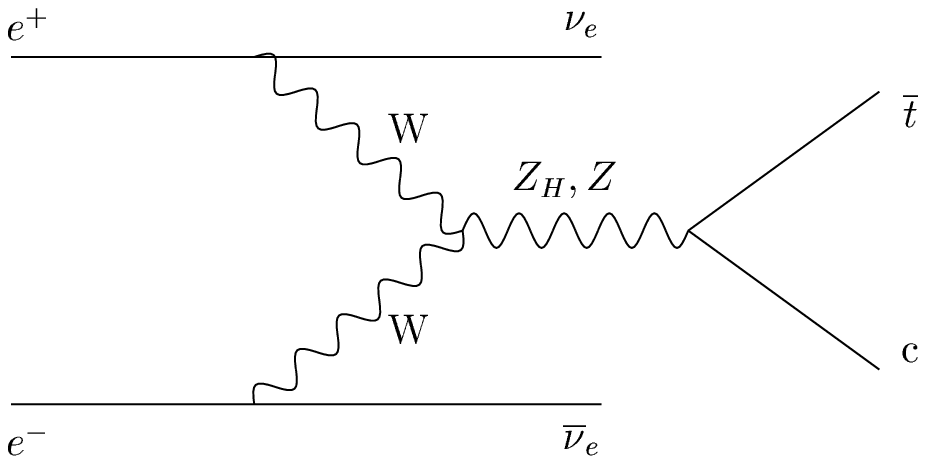,width=460pt,height=700pt} \vspace{-12cm}
\hspace{5mm} \caption{Feynman diagrams contribute to the
$WW$-fusion process $e^{+}e^{-}\rightarrow WW
\nu_{e}\overline{\nu}_{e}\rightarrow\hspace*{1.8cm}\overline{t}c\nu_{e}
\overline{\nu_{e}}$.} \label{ee}
\end{center}
\end{figure}

The process $e^{+}e^{-}\rightarrow(\overline{t}c +
t\overline{c})\nu_{e}\overline{\nu}_{e}$ can be well approximated
by the $WW$-fusion process
$W^{+}_{\lambda_{+}}W^{-}_{\lambda_{-}}\rightarrow \overline{t}c +
t\overline{c}$. It has been shown the effective $W$-boson
approximation $(EWA)$ provides a viable simplification for
$WW$-fusion processes at the high $CM$ energies[13]. Thus, we use
the effective $EWA$ to estimate the production cross section of
the process $e^{+}e^{-}\rightarrow W^{+}W^{-}
\nu_{e}\overline{\nu}_{e}\rightarrow(\overline{t}c+t\overline{c})
\nu_{e}\overline{\nu_{e}}$ in the future $LC$ experiments with
$\sqrt{S}\geq1TeV$.

For the subprocess $W^{+}_{\lambda_{+}}W
^{-}_{\lambda_{-}}\rightarrow\overline{t}c+t\overline{c}$
generated by the gauge bosons $Z_{H}$ and $Z$ with the helicities
$\lambda_{\pm}=0, \pm1$, the non-vanishing helicity amplitudes are
$M_{+1+1}=M_{-1-1}$, $M_{+10}=M_{-10}$ and $M_{00}$[6]. The
production cross section $\widehat{\sigma}(\widehat{s})$ of this
subprocess contributed by $Z$ exchange and $Z_{H}$ exchange can be
written as:
\begin{eqnarray}
\widehat{\sigma}(\widehat{s})\nonumber&=&\widehat{\sigma}_{11}(\widehat{s})+
\widehat{\sigma}_{-1-1}(\widehat{s})
+\widehat{\sigma}_{10}(\widehat{s})+\widehat{\sigma}_{-10}(\widehat{s})
+\widehat{\sigma}_{00}(\widehat{s})\\
&=&(A_{1}+A_{2}+A_{3})[1+1+\frac{\widehat{s}}{2M^{2}_{W}}+\frac{\widehat{s}}{2M^{2}_{W}}
+(1+\frac{\widehat{s}}{2M^{2}_{W}})^{2}]
\end{eqnarray}
with
\begin{eqnarray}
&&A_{1}=\frac{32\pi\alpha^{2}_{e}K^{2}_{tc}}{3S^{4}_{W}}\beta^{4}_{t}\beta_{W}
(1+\frac{m^{2}_{t}}{2\widehat{s}})\widehat{s}\chi^{2}_{Z},\\
&&A_{2}=\frac{8\pi\alpha^{2}_{e}K^{2}_{tc}}{3S^{4}_{W}C^{2}_{W}}\frac{\nu^{4}}{f^{4}}
[c^{4}(c^{2}-s^{2})^{2}]\beta^{4}_{t}\beta_{W}(1+\frac{m^{2}_{t}}{2\widehat{s}})
\widehat{s}\chi^{2}_{Z_{H}},\\
&&A_{3}=-\frac{32\pi\alpha^{2}_{e}K^{2}_{tc}}{3S^{4}_{W}C^{2}_{W}}\frac{\nu^{2}}{f^{2}}
[c^{2}(c^{2}-s^{2})]\beta^{4}_{t}\beta_{W}(1+\frac{m^{2}_{t}}{2\widehat{s}})
\cdot\widehat{s}R_{e}\mid\chi_{Z}\cdot\chi_{Z_{H}}\mid,
\end{eqnarray}
where $\sqrt{\widehat{s}}$ is the $CM$ energy of the subprocess
$W^{+}_{\lambda_{+}}W^{-}_{\lambda_{-}}\rightarrow\overline{t}c+t\overline{c}$,
$\beta_{t}=\sqrt{1-\frac{m^{2}_{t}}{\widehat{s}}}$, and
$\beta_{W}=\sqrt{1-\frac{4M^{2}_{W}}{\widehat{s}}}$. The factors
$A_{1}$, $A_{2}$ and $A_{3}$ come from $Z$ exchange, $Z_{H}$
exchange, and interference between $Z$ and $Z_{H}$, respectively.

In general, the cross section $\sigma(\overline{t}c)$ for the
process $e^{+}e^{-}\rightarrow W^{+}W^{-}
\nu_{e}\overline{\nu}_{e}\rightarrow(\overline{t}c +
t\overline{c})\nu_{e}\overline{\nu}_{e}$ can be obtained by
folding the cross section $\widehat{\sigma}(\widehat{s})$ for the
subprocess $W^{+}_{\lambda_{+}}W^{-}_{\lambda_{-}}\rightarrow
\overline{t}c + t\overline{c}$ with $W^{\pm}_{\lambda_{\pm}}$
distribution functions $f^{W^{\pm}}_{\lambda_{\pm}}$:
\begin{eqnarray}
\sigma(\overline{t}c)\nonumber
&=&\sum_{\lambda_{+}\lambda_{-}}\int^{1}_{m_{t}/\sqrt{s}}2xdx\int^{1}_{x^{2}}
\frac{dx_{+}}{x_{+}}f^{W^{+}}_{\lambda_{+}}(x_{+})f^{W^{-}}_{\lambda_{-}}
(\frac{x^{2}}{x_{+}})\widehat{\sigma}(W^{+}_{\lambda_{+}}W^{-}_{\lambda_{-}}
\rightarrow \overline{t}c + t\overline{c})\\\nonumber
&=&\int^{1}_{m_{t}/\sqrt{s}}2xdx\int^{1}_{x^{2}}\frac{dx_{f}}{x_{+}}[f^{W}_{+}(x_{+})
f^{W}_{+}(\frac{x^{2}}{x_{+}})+f^{W}_{-}(x_{+})f^{W}_{-}(\frac{x^{2}}{x_{+}})+f^{W}_{-}(x_{+})
f^{W}_{0}(\frac{x^{2}}{x_{+}})\frac{\widehat{s}^{2}}{2m^{2}_{W}}\\
&&+f^{W}_{+}(x_{+})f^{W}_{0}(\frac{x^{2}}{x_{+}})\frac{\widehat{s}^{2}}{2m^{2}_{W}}+
f^{W}_{0}(x_{+})f^{W}_{0}(\frac{x^{2}}{x_{+}})(1+\frac{\widehat{s}^{2}}{2m^{2}_{W}})^{2}]
(A_{1}+A_{2}+A_{3})
\end{eqnarray}
In our calculation, we will use the full distribution functions
$f^{W^{\pm}}_{\lambda_{\pm}}(x)$ given by Refs[13,14] and
$\widehat{s}=x^{2}S$.

The production cross section $\sigma(\overline{t}c)$ for the
process $e^{+}e^{-}\rightarrow W^{+}W^{-}
\nu_{e}\overline{\nu}_{e}\rightarrow(\overline{t}c +
t\overline{c})\nu_{e}\overline{\nu}_{e}$ is plotted in Fig.3 as a
function of the $CM$ energy $\sqrt{S}$ for the free parameters
$c=0.4$ and $M_{Z_{H}}=1.5TeV$. Our numerical results show that
the production cross section $\sigma(\overline{t}c)$ mainly comes
from $Z$ exchange and is not sensitive to the free parameters $c$
and $M_{Z_{H}}$. In most of the parameter space preferred by the
electroweak precision data, the value of $\sigma(\overline{t}c)$
is about $1.5\times10^{-3}fb$. Thus, the possible $FC$ signals of
the $LH$ model are very difficult to be detected via the process
$e^{+}e^{-}\rightarrow W^{+}W^{-}
\nu_{e}\overline{\nu_{e}}\rightarrow(\overline{t}c +
t\overline{c})\nu_{e}\overline{\nu}_{e}$ in future $LC$
experiments.

\begin{figure}[htb]
\vspace{-0.5cm}
\begin{center}
\epsfig{file=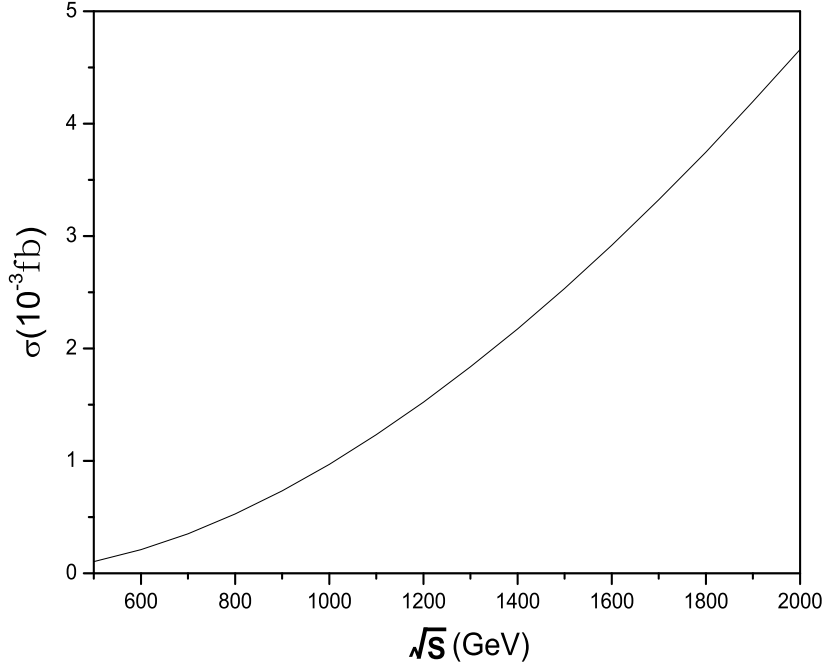,width=350pt,height=300pt} \vspace{-1.0cm}
\hspace{5mm} \caption{The production cross section
$\sigma(\overline{t}c)$ as a function of the $CM$ energy
$\sqrt{S}$ for \hspace*{1.8cm} $M_{Z_{H}}=1.5TeV$ and $c=0.4$.}
\label{ee}
\end{center}
\end{figure}

Certainly, the $FC$ couplings $Z\overline{t}c$ and
$Z_{H}\overline{t}c$ can also has contribute to the top-charm
production at the $LC$ experiments via the process
$e^{+}e^{-}\rightarrow
ZZe^{+}e^{-}\rightarrow\overline{t}ce^{+}e^{-}$. The main
difference between $\sigma(\overline{t}c)$ and
$\sigma(\overline{t}ce^{+}e^{-})$ arises from the dissimilarity
between the distribution functions for $W$ and $Z$ bosons. Since
the $W$ distribution function is larger than $Z$ distribution
function, $\sigma(\overline{t}ce^{+}e^{-})$ is smaller than
$\sigma(\overline{t}c)$ by about one order of magnitude[5]. Thus,
we do not need to further consider the process
$e^{+}e^{-}\rightarrow\overline{t}ce^{+}e^{-}$ in the context of
the $LH$ model.

\noindent{\bf 4. The process $e^{-}\gamma\rightarrow e^{-}
\overline{t}c$ in the $LH$ model }

A future $LC$ can also operate in $e^{-}\gamma$ collisions, where
the $\gamma$-beam is generated by the backward Compton scattering
of the incident positron- and laser-beam. Its energy and
luminosity can reach the same order of magnitude of the
corresponding positron beam[15]. From Eq.(1), we can see that the
$FC$ couplings $Z\overline{t}c$ and $Z_{H}\overline{t}c$ might
have significant contributions to the top-charm production via the
process $e^{-}\gamma\rightarrow e^{-}Z^{*}\gamma\rightarrow e^{-}
\overline{t}c$. The relevant Feynman diagrams are shown in Fig.4.

\begin{figure}[htb]
\vspace{-8.0cm}
\begin{center}
\epsfig{file=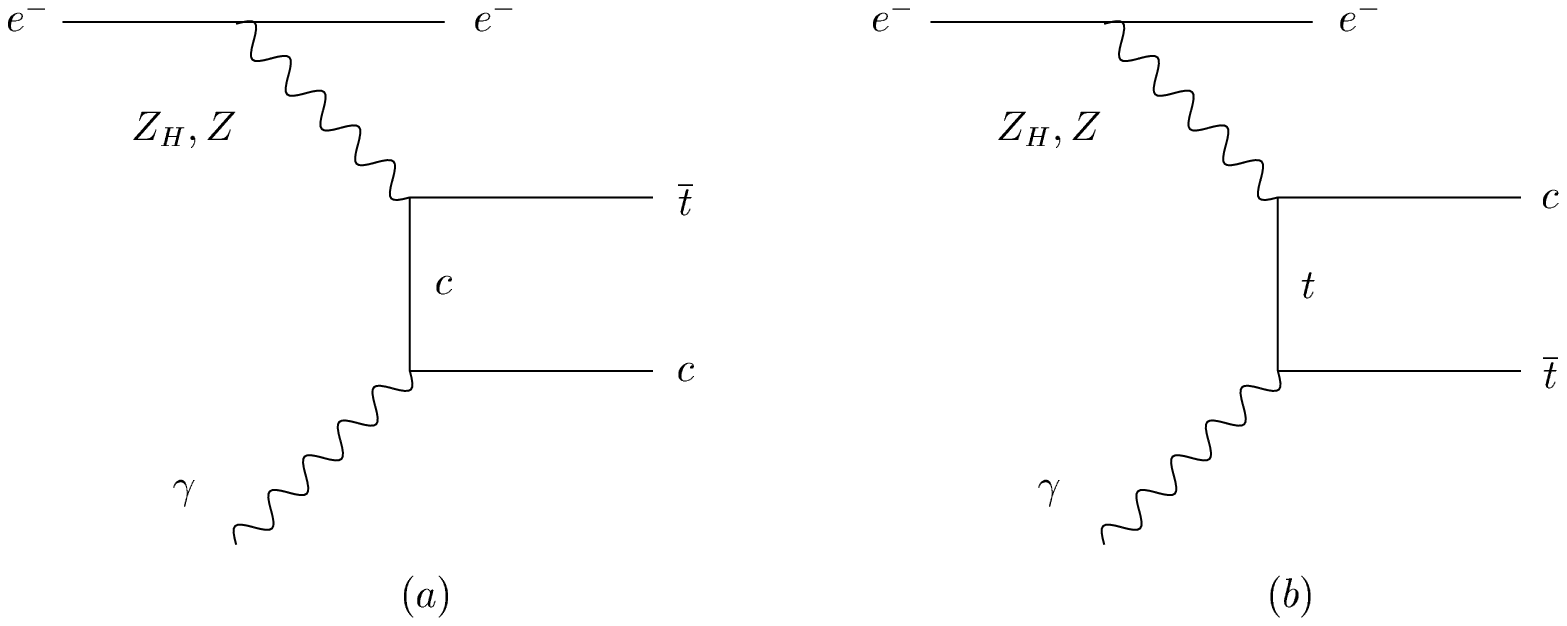,width=460pt,height=680pt} \vspace{-11cm}
\hspace{5mm} \caption{Feynman diagrams contribute to the process
$e^{-}\gamma\rightarrow e^{-}\overline{t}c$} \label{ee}
\end{center}
\end{figure}

For the subprocess $Z(P_{Z})+\gamma(k)\rightarrow
t(P_{t})+\overline{c}(P_{c})$, we define the kinematical invariant
$t=(P_{t}-P_{Z})^{2}$. The renomalization amplitude can be written
as:
\begin{eqnarray}
M\nonumber &=&M_{Z}+M_{Z_{H}}\\
\nonumber&=&-\frac{e^{2}}{3S_{W}C_{W}}K_{tc}\overline{u}_{c}\gamma_{\mu}\frac{i}{
\not P_{Z}-\not
P_{t}-m_{c}+i\epsilon}\gamma_{\nu}P_{L}v_{t}\epsilon^{\mu}(k)
\epsilon^{\nu}(Z)\\\nonumber
&&-\frac{e^{2}}{3S_{W}C_{W}}K_{tc}v_{t}\gamma_{\mu}\frac{i}{\not
P_{Z}-\not
P_{c}-m_{t}+i\epsilon}\gamma_{\nu}P_{L}\overline{u}_{c}\epsilon^{\mu}(k)
\epsilon^{\nu}(Z)\\\nonumber
&&-\frac{e^{2}}{3S_{W}C_{W}}\frac{c}{s}K_{tc}\overline{u}_{c}\gamma_{\mu}\frac{i}{\not
P_{Z_{H}}-\not
P_{t}-m_{c}+i\epsilon}\gamma_{\nu}P_{L}v_{t}\epsilon^{\mu}(k)
\epsilon^{\nu}(Z_{H})\\
&&-\frac{e^{2}}{3S_{W}C_{W}}\frac{c}{s}K_{tc}v_{t}\gamma_{\mu}\frac{i}{\not
P_{Z_{H}}-\not
P_{c}-m_{t}+i\epsilon}\gamma_{\nu}P_{L}\overline{u}_{c}\epsilon^{\mu}(k)
\epsilon^{\nu}(Z_{H})
\end{eqnarray}
with $P_{L}=\frac{1-\gamma_{5}}{2}$.

The effective cross section $\sigma(e^{-}\overline{t}c)$ at a $LC$
with $CM$ energy $\sqrt{S}$ can be obtained by folding the
subprocess cross section
$\widehat{\sigma}(Z\gamma\rightarrow\overline{t}c)$ with the gauge
boson $Z$ and photon distribution functions $f_{Z/e}$[13,14] and
$f_{\gamma/e}$[16]:
\begin{equation}
\sigma(e^{-}\overline{t}c)=\int^{0.83}_{(m_{t}+m_{c})^{2}/S}d\tau
\int^{1}_{\tau/0.83}\frac{dx}{x}f_{\gamma/e}(\frac{\tau}{x})f_{Z/e}(x)
\widehat{\sigma}(Z\gamma\rightarrow\overline{t}c)
\end{equation}
In above equation, we have assumed $\widehat{s}=\tau S$, in which
$\widehat{s}$ is the $CM$ energy of the subprocess
$Z\gamma\rightarrow\overline{t}c$.

\begin{figure}[htb]
\vspace{-0.5cm}
\begin{center}
\epsfig{file=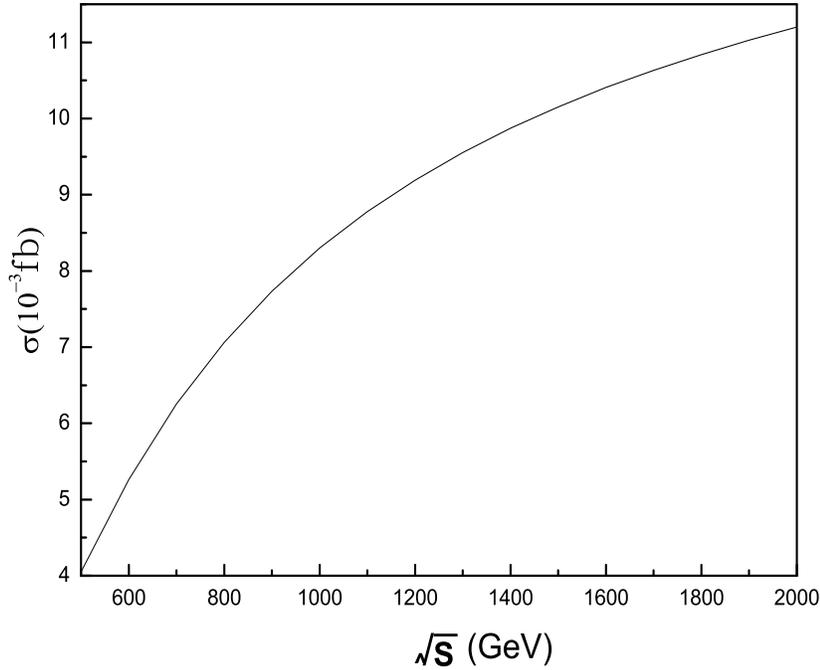,width=350pt,height=300pt} \vspace{-1.0cm}
\hspace{5mm} \caption{The cross section
$\sigma(e^{-}\overline{t}c)$ as a function of the $CM$ energy
$\sqrt{S}$ for $c=0.4$ and \hspace*{1.8cm} $M_{Z_{H}}=2TeV$.}
\label{ee}
\end{center}

\end{figure}

Observably, the contributions of the $LH$ model to the process
$e^{-}\gamma\rightarrow e^{-}\overline{t}c$ mainly come from the
$FC$ coupling $Z\overline{t}c$. The cross section
$\sigma(e^{-}\overline{t}c)$ is not sensitive to the mixing
parameter $c$ and the $Z_{H}$ mass $M_{Z_{H}}$. Thus, in our
numerical estimation, we taken $c=0.4$ and $M_{Z_{H}}=2TeV$. Our
numerical result are shown in Fig.5. One can see from Fig.5 that
the cross-section $\sigma(e^{-}\overline{t}c)$ increases as the
$CM$ energy $\sqrt{S}$ increasing. For $\sqrt{S}\leq2TeV$, the
value of $\sigma(e^{-}\overline{t}c)$ is smaller than
$1.2\times10^{-2}fb$. Thus, the possible $FC$ signals of the $LH$
model is very difficult to be detected via the process
$e^{-}\gamma\rightarrow e^{-}\overline{t}c$ in  future $LC$
experiments.

\noindent{\bf 5. Conclusions}

Little Higgs theory has generated much interest as one kind of
models of $EWSB$, which can be regarded as one of the important
candidates of new physics beyond the $SM$. For all of the little
Higgs models, at least one vector-like top quark $T$ is needed to
cancel the numerically most large quadratic divergence coming from
top Yukawa couplings. Due to the presence of extra quarks, the
$CKM$ matrix is not unitary and $FCNC's$ may exist at tree-level.
Thus, the little Higgs models generally predict the $FC$ couplings
$Z\overline{t}c$ and $Z_{H}\overline{t}c$.

In this Letter, we study the contributions of the $FC$ couplings
predicted by the $LH$ model to the top charm production via the
the processes $e^{+}e^{-}\rightarrow\overline{t}c +t\overline{c}$,
$e^{+}e^{-}\rightarrow(\overline{t}c
+t\overline{c})\nu\overline{\nu_{e}}$, and $e^{-}\gamma\rightarrow
e^{-}\overline{t}c$ in future $LC$ experiments. We find that the
cross sections of the processes $e^{+}e^{-}\rightarrow (
\overline{t}c +t\overline{c})\nu\overline{\nu_{e}}$ and
$e^{-}\gamma\rightarrow e^{-}\overline{t}c$ are very small in all
of the parameter space, which can not give detectable signals. For
the process $e^{+}e^{-}\rightarrow\overline{t}c +t\overline{c}$,
the top-charm production cross section is approximately equal to
$1.2\times10^{-2}fb$ in part of the parameter space preferred by
the electroweak precision data. However, for $\sqrt{S}\approx
M_{Z_{H}}$, the cross section $\sigma(S)$ can be significantly
enhanced and the contributions of the $LH$ model to the top-charm
production mainly come from the $FC$ couplings
$Z_{H}\overline{t}c$. For example, for $\sqrt{S} \approx
M_{Z_{H}}=1TeV$, $1.5TeV$, $2TeV$, the value of $\sigma(S)$ is
$16.8fb$, $7.7fb$, and $4.4fb$, respectively. The resonance
effects of the heavy gauge boson $Z_{H}$ on the top-charm
production should be detected in future $LC$ experiments.

\vspace{0.5cm} \noindent{\bf Acknowledgments}

This work was supported in part by Program for New Century
Excellent Talents in University(NCET-04-0290), the National
Natural Science Foundation of China under the grant No.90203005
and No.10475037, and the Natural Science Foundation of the
Liaoning Scientific Committee(20032101).

\null
\end{document}